\documentclass[%
reprint,
superscriptaddress,
 amsmath,amssymb,
 aps,
 prl,
]{revtex4-1}

\usepackage{textcomp}
\usepackage{placeins}

\usepackage{placeins}
\usepackage{graphicx}

\usepackage{epstopdf}

\usepackage[dvipsnames]{xcolor}

\usepackage{dcolumn}
\usepackage{bm}

\usepackage{siunitx}

\usepackage{hyperref}
\hypersetup{colorlinks,linkcolor=blue,citecolor=blue}
\begin{document}

\preprint{APS/123-QED}

\title{Probing the spatial distribution of k-vectors in situ with Bose-Einstein condensates}

\author{Samuel Gaudout}
\author{Rayan Si-Ahmed}
\author{Cl\'{e}ment Debavelaere}
\author{Menno Door}
\author{Pierre Clad\'{e}}
\affiliation{Laboratoire Kastler Brossel, Sorbonne Universit{\'{e}}, CNRS, ENS-Universit{\'{e}} PSL, Coll{\`{e}}ge de France, 75005 Paris, France}
\author{Saïda Guellati-Khelifa}
\email{guellati@lkb.upmc.fr}
\affiliation{Laboratoire Kastler Brossel, Sorbonne Universit{\'{e}}, CNRS, ENS-Universit{\'{e}} PSL, Coll{\`{e}}ge de France, 75005 Paris, France}
\affiliation{Conservatoire National des Arts et M\'{e}tiers, 75003 Paris, France}

\date{\today}

\begin{abstract}

We present a novel method for mapping \textit{in situ} the spatial distribution of photon momentum across a laser beam using a Bose-Einstein condensate (BEC) as a moving probe. By displacing the BEC, we measure the photon recoil by atom interferometry at different positions in the laser beam and thus reconstruct a two-dimensional map of the local intensity and effective dispersion of the $\vec{k}$ wave vector. Applied to a beam diffracted by a diaphragm, this method reveals a local \textit{extra recoil} effect, which exceeds the magnitude $h\nu/c$ of the individual plane waves over which the beam can be decomposed. This method offers a new way to precisely characterize wavefront distortions and to evaluate one of the major systematic bias sources in quantum sensors based on atom interferometry.
\end{abstract}

\maketitle

When an atom absorbs or emits a photon, it experiences a recoil due to the momentum carried by the photon, given by $\vec{p}=\hbar \vec{k}$, where $\vec{k}$ is the photon wave vector. This exchange of photon momentum with atoms is a key aspect of atom-light interactions, playing a crucial role in a wide range of applications. 
For a plane wave, the direction of $\vec{k}$ is well defined, as is its magnitude, given by $k_0 = 2\pi \nu_0 / c$, where $\nu_0$ denotes the optical frequency. Deviations from this ideal case lead to spatial variations in both the direction and magnitude of $\vec{k}$. An interesting effect is that the local magnitude of $\vec{k}$ may exceed the nominal value of each plane-wave component forming the optical beam~\cite{berry1994,barnett2013, MATSUDO199864}. This phenomenon, hereafter referred to as \textit{extra recoil}, was experimentally observed by exploiting the correlation between photon recoil and intensity in a distorted beam \cite{Bade2018}.

In light-pulse atom interferometry, accurate knowledge and precise control of the photon momenta transferred to the atoms are crucial. Such interferometers operate by coherently splitting and recombining atomic wave packets using sequences of laser pulses~\cite{kasevich_atomic_1991, mueller_atom_2008}. Each interaction imparts a momentum kick of $\hbar \vec{k}$, creating distinct momentum states that follow separate trajectories within the interferometer. The magnitude of the resulting phase shift between the interfering paths depends directly on the value of $\vec{k}$. Even small distortions in the optical wavefront have a significant impact on the performance of atom interferometers, notably for advanced applications in quantum metrology, gravitational wave detection, and measurement of fundamental constants \cite{morel_determination_2020, parker2018, Schubert2024, Overstreet2022, Overstreet2017, Asenbaum2020, Gautier2022, 
Savoie2018, Gebbe2021}. In recent years, efforts have focused on two main directions. One focuses on designing high-quality optical beams using advanced techniques such as beam shaping with high-quality optics or adaptive optics, which all aim to minimize distortions and ensure uniform wavefronts~\cite{hamilton_atom_2015, trimeche_active_2017, Muntinga2013}. The other aims at investigating methods to measure the beam profile as seen by the atoms ~\cite{Fils2005, gauguet_characterization_2009, schkolnik_effect_2015, zhou_observing_2016, karcher_improving_2018, xu_situ_2024, seckmeyer_spatially_2025, Pagot2025}. 

In this Letter, we present a novel method for reconstructing, \textit{in situ}, a 2D map of the spatial distribution of wave vectors. The method relies on measuring photon recoil at selected points across the transverse profile of a laser beam using an atom interferometer. To probe this recoil locally, we use a Bose-Einstein condensate (BEC), whose size remains, after propagation, much smaller than that of an optical molasses. To perform the mapping, we developed a technique to move and precisely control the position of the BEC, providing access to the distribution of wavevectors as experienced by the atoms inside the vacuum chamber. This approach also enables an \textit{in situ} measurement of the beam intensity, a crucial parameter for modeling \(\vec{k}\). Furthermore, the use of a BEC offers direct access to the \textit{extra recoil} effect. This effect was amplified by deliberately shaping the beam profile with an iris.

\begin{figure*}
    \centering
    \includegraphics[width=\linewidth]{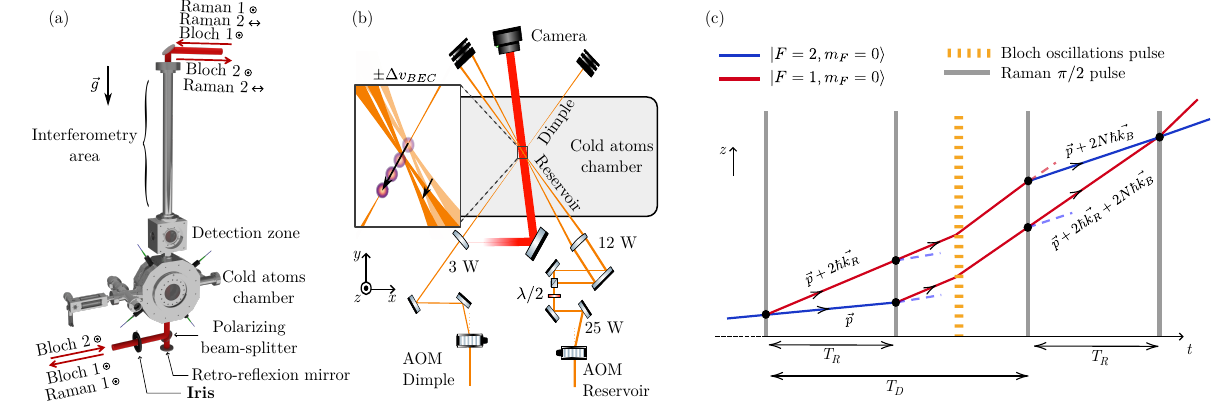} 
    \caption{(\textit{a}) Schematic of the experimental setup showing the cold atoms chamber, the detection and the atom interferometry areas, as well as the propagation of the Raman and Bloch beams along the vertical axis ($z$-axis). Raman beams are in a orthogonal linear polarizations arrangement ($\bigodot$ and $\leftrightarrow$) while both Bloch beams have the same linear polarization ($\bigodot$). (\textit{b}) Experimental setup used to generate the laser beams forming the dipole trap for the BEC production. We employ two acousto-optic modulators (AOMs) to control the position of the center of the dipole trap and to impart a transverse velocity to the Bose-Einstein condensate before being released. The inset illustrates how a change of the trap position is used to impart a velocity to the BEC  (\textit{c}) Atom trajectories (for simplicity in the free falling reference frame) and temporal sequence used to measure recoil velocity using a Ramsey-Bordé interferometer. It consists of two pairs of $\pi/2$ pulses with $T_R=20$ ms, separated by a duration $T_D=35$ ms. These pulses induce stimulated Raman transitions between the two hyperfine states \(|F=1\rangle\) and \(|F=2\rangle\) of 87 rubidium atom. They separate and recombine the atomic wave packets that interfere at the end of the time sequence. Between the two pairs of pulses, an accelerated optical lattice is switched on for 6~ms. Atoms perform $N_B=500$ Bloch oscillations and acquire a momentum of $2 N_\mathrm{B} \hbar \vec{k}_\mathrm{B}$.} 
    \label{fig:schemas_bec_movement} 
\end{figure*}

In order to understand how and why the wave vector may vary depending on the light field properties, we consider a laser beam with frequency $\nu_0$ and phase $\phi(\vec{r})$ propagating primarily along an axis $\vec{u}_z$. The momentum of a photon is given by the canonical momentum \(\vec{p} = \hbar \vec{k}=\hbar\vec{\nabla} \phi\)~\cite{berry2013,Antognozzi}. We define $\vec{\kappa} = \vec{k}/k_0 - \vec{u}_z$ the relative variations of the wave vector defined such that the \(z\) component \(p_z\) of the momentum at position \(z_0\) is $p_z = \hbar k_0 (1 + \kappa_z)$. In a previous work \cite{Bade2018} we evaluated the correction $\kappa_z$ under the paraxial approximation and obtained :
\begin{equation}
\kappa_z = - \frac{1}{2k_0^2}\left|\left|\vec{\nabla}_{\perp}\phi(\vec{r})\right|\right|^2 + \frac{1}{4k_0^2}\frac{\Delta_{\perp}I(\vec{r})}{I(\vec{r})},
\label{eq:delta_keff}
\end{equation}
where the gradient operator $\vec\nabla_\perp$ and the Laplacian operator $\Delta_\perp$ are evaluated in the plane $z = z_0$.  As discussed in \cite{Bade2018}, the first term, associated with the phase gradient, corresponds to a tilt with respect to the propagation direction, resulting from a local distortion of the wavefront. The second term represents a correction for the momentum induced by spatial variations in intensity. The relative fluctuations of intensity and the phase fluctuations have the same standard deviation $\sigma$, because wavefront distortions induced by imperfections in the optical system are converted into intensity fluctuations during laser beam propagation. Consequently, the contribution of the phase gradient becomes negligible compared to that of the Laplacian intensity, since it scales as $\sigma^2$. When the atomic cloud that probes the recoil is too large - as is typically the case with optical molasses - the Laplacian of the field is averaged over the spatial extent of the cloud, which reduces the amplitude of the effect observed due to local variations in the optical field. In Ref.~\cite{Bade2018}, the contribution of the Laplacian term was observed using an optical molasses, by exploiting correlations between photon recoil and the efficiency of Bloch oscillations, which in turn depend on the laser intensity. 

\begin{figure*}
\includegraphics[width=\linewidth]{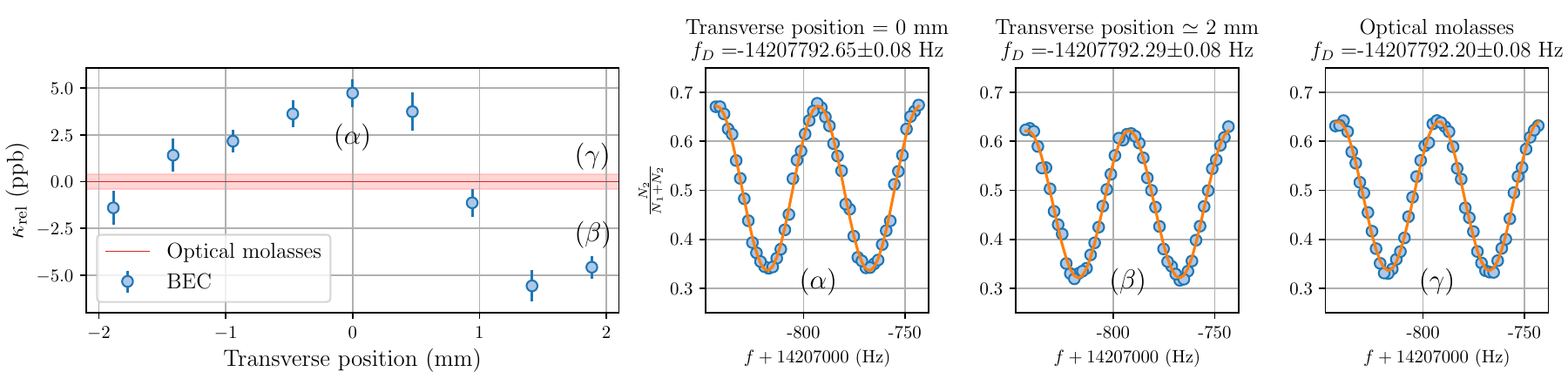}
\caption{Typical interference fringes shown for different configurations : $(\alpha)$ BEC positioned at the beam center, $(\beta)$ BEC with transverse velocity displaced by $\sim$ 2 mm from the beam center and $(\gamma)$  measurements performed with optical molasses. Right) Values of \(\kappa_\mathrm{rel}\) extracted from Eq.~\ref{Kappa} using measurements of the Doppler shifted frequency \(f_D\) and as reference frequency $f_D^\mathrm{ref}$ obtained with optical molasses. The data were acquired over 117 hours, using BEC and optical molasses alternately. }
\label{Fig_flat_Beams}
\end{figure*}

The experimental setup is shown in Fig.~\ref{fig:schemas_bec_movement}. The BEC is produced by evaporative cooling in an optical dipole trap. After preparation, it is released and then launched vertically using an accelerated optical lattice. Atom-interferometric measurements are performed after a 190~ms time of flight, approximately one meter above the trap. To probe the intensity profile of the laser beam, the BEC is displaced transversely by imparting an initial horizontal velocity at the end of the evaporation process, this velocity being chosen so that the condensate reaches the desired position at the time of measurement. The dipole trap (see Fig.~\ref{fig:schemas_bec_movement}-b) consists of the intersection of three laser beams \cite{Zhibin2022}. To impart transverse velocity, we apply a frequency shift to the acousto-optic modulators (AOMs) controlling the trapping beams (Fig.~\ref{fig:schemas_bec_movement}-b). This shifts the center of the trap, accelerating the BEC. The AOMs are turned off after a quarter oscillation of the cloud within the dipole trap when the atoms reach their maximum transverse velocity. This velocity is proportional to the trap displacement — and thus to the AOM frequency shift — and can be precisely controlled. For calibration, the BEC trajectory is tracked via absorption imaging using two cameras: one horizontal and another (not shown) tilted 30$^\circ$ relative to the vertical axis. The maximum transverse velocity we impart to the BEC is \SI{10}{\milli\meter\per\second} along the $x$ and $y$ axes, with an RMS fluctuation of $\pm \SI{0.075}{\milli\meter\per\second}$, negligible compared to the cloud's velocity spread of $\sim \SI{1.8}{\milli\meter\per\second}$. The maximum displacement of the cloud when the measurement is performed is nearly \SI{2}{\milli\meter} and its size is \SI{350}{\micro\meter}.

For the direct measurement of local photon recoil, we employ a Ramsey-Bordé atom interferometer combined with Bloch oscillations \cite{cadoret_combination_2008}. As illustrated in Fig.~\ref{fig:schemas_bec_movement}-c), the sequence comprises two pairs of Raman \(\pi/2\) pulses. Each Raman pulse consists of two counter-propagating laser beams with wave vectors $\vec{k}_{R1}$ and $\vec{k}_{R2}$, driving a coherent two-photon transition between the $^{87}$Rb hyperfine states \(|F=1\rangle\) and \(|F=2\rangle\). The Raman transition acts as an atomic beamsplitter by creating a superposition of two separate atomic wavepackets. Between the pulse pairs, an accelerated optical lattice—formed by another counter-propagating beam pair with wave vectors \(\vec{k}_{B1}\) and \(\vec{k}_{B2}\), where \(|\vec{k}_{B1}| \simeq |\vec{k}_{B2}| = k_B\)—induces Bloch oscillations, transferring momentum \(N_B \hbar (\vec{k}_{B1} - \vec{k}_{B2})\) to the atoms. Atomic interference fringes are obtained by measuring the populations in \(|F=1\rangle\) and \(|F=2\rangle\) while scanning the Raman frequency of the second pulse pair. The central fringe position indicates the Raman frequency shift that compensates for the Doppler effect associated with the \(2 N_B\) photon momentum transfer. To cancel the effect of constant gravity and other level shifts, measurements are repeated using four configurations: alternating the directions of both the Bloch acceleration and the Raman wave vectors, as described in Refs.~\cite{bouchendira_new_2011,morel_determination_2020}. This procedure yields a precise determination of the Doppler frequency given by:
\begin{equation}
    2\pi f_D = \frac{N_\text{B} \hbar\, \left(\vec{k}_\mathrm{B2}-\vec{k}_\mathrm{B1}\right) \cdot \left(\vec{k}_\mathrm{R2}-\vec{k}_\mathrm{R1}\right)}{m}\,.
\end{equation}

This measurement procedure is usually used to determine the $h/m$ ratio of the atom. Here, we reverse the approach and use it to extract $\vec{\kappa}$, the relative variations of the four wave vectors (from both Raman and Bloch beams), due to their deviations from the main propagation axis~: 
\begin{equation}
   2 \pi f_D \simeq \frac{4 N_\text{B} \hbar\, k_\mathrm{B} k_\mathrm{R}}{m} \left(1+\frac{1}{2}\vec{\kappa} \cdot \vec{u}_z\right)
   \label{Kappa}
\end{equation}
where $\vec{\kappa}=\left(\vec{\kappa}_\mathrm{B2}-\vec{\kappa}_\mathrm{B1}+\vec{\kappa}_\mathrm{R2}-\vec{\kappa}_\mathrm{R1}\right)$ and \(k_R =\left(|\vec{k}_\mathrm{R1}| + |\vec{k}_\mathrm{R2}|\right)/2\)

Fig.~\ref{Fig_flat_Beams} shows measurements performed by displacing the BEC along a horizontal line in the transverse plane of the Raman and Bloch laser beams. This line spans positions ranging from $-2$~mm to $+2$~mm relative to the center of the beam, with a beam waist of 5~mm. It should be noted that the position we are referring to is the position of the atomic cloud at the moment when we turn on the optical lattice to transfer $2 N_\mathrm{B}$ photon momenta. Since the optical lattice is turned on for 6 ms, during this time the cloud moves by a maximum of 60 $\mu$m. We predefine nine discrete positions along this line, each corresponding to specific driving frequencies of the AOMs controlling the dipole trap. Values of \( \kappa_z \) are extracted from Doppler shift measurements \(f_D\) using Eq.~\ref{Kappa}. One value of \(f_D\) is extracted from four spectra recorded in 15 minutes, corresponding to the four configurations described previously. Figure~\ref{Fig_flat_Beams} shows typical spectra for the same configuration, obtained with the BEC at the center ($\alpha$), shifted by approximately 2 mm ($\beta$), and with optical molasses ($\gamma$). At around 2 mm, we lose about 10$\%$  of the contrast, but the uncertainty on $f_D$ remains unchanged at approximately 80 mHz, which corresponds to a relative uncertainty of $2.8 \times 10^{-9}$. Each data point in Fig.~\ref{Fig_flat_Beams}-left represents an average over roughly 18 such measurements, yielding a statistical uncertainty of $6 \times 10^{-10}$ on \(\kappa\). The effects of local intensity fluctuations are averaged when using optical molasses. In order to extract a local variation in $\kappa$ using only Doppler frequency measurements with a BEC, we use a measurement with molasses, $f_D^\mathrm{ref}$, as a reference. To obtain $f_D^\mathrm{ref}$, we average 53 values, their relative standard deviation is $3.8\times 10^{-10}$ with $\chi^2=1.2$. Data are acquired alternating BECs and optical molasses. The quantity $\kappa_\mathrm{rel}$ shown in the figure is obtained from $f_D/f_D^\mathrm{ref} - 1$.

The data show spatial variations of \(\kappa\) (or equivalently \( f_D \)) as the BEC is displaced. In the ideal case of a perfect Gaussian beam, \(\kappa\) can be computed as a function of the radial coordinate \( r \). At the beam waist, where the wavefronts are almost planar, the expected value of \(\kappa_z\), which can be derived from the analytical phase profile, follows the expression \(\kappa_z = -\frac{2}{k^2 w^2} \left(1 - \frac{r^2}{w^2}\right)\). This result corresponds to the Gouy phase correction~\cite{Wicht_2002,Wicht_2005,CladePRA2006}. For displacements in the range \( r = 0 \) to \( 2~\mathrm{mm} \), the corresponding variation in \(\kappa\) is of the order of \( \num{2e-10} \), which is negligible compared to the much larger deviations observed experimentally. This suggests the presence of significant wavefront distortions beyond those expected from a simple Gaussian model. We also took a measurement using a 2D map on 121 positions centered around our beam. The average value of these measurements is in good agreement with the values obtained using an optical molasses instead of a condensate.

\begin{figure}
\includegraphics[width=0.99\columnwidth]{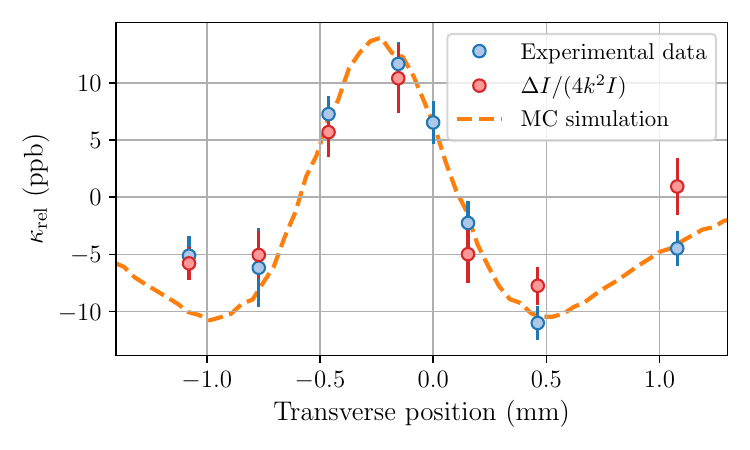}
\caption{
One-dimensional profile of $\kappa_{\mathrm{rel}}$ measured with a BEC along the $y$-axis are depicted by blue dots, using a beam clipped by a diaphragm. The $\kappa_{\mathrm{rel}}$ correction calculated from the beam intensity measurement is illustrated with red dots. These measurements are compared to Monte-Carlo simulations of the experiment, represented by an orange dashed line.}
\label{Fig_donuts_1D}
\end{figure}

To further test our measurement technique, we deliberately introduced strong spatial aberrations in the bottom Bloch beam by inserting a 4 mm-diameter iris at the collimator output, as presented in Fig. \ref{fig:schemas_bec_movement}. After a propagation distance of 2.7~m, this hard aperture generates an intensity and phase pattern at the atoms' location. The iris size and position were carefully chosen so that the center of the beam, where the atoms are initially located, corresponds to an intensity minimum.  We then performed a position-resolved measurement of the atomic recoil velocity. The resulting values of $\kappa_\mathrm{rel}$  are shown in Fig.~\ref{Fig_donuts_1D}. We see that the recoil correction reverses sign as the condensate is moved from the dark central region to the first bright ring. This inversion of $\kappa_z$  is a direct manifestation of the Laplacian term in Eq.~\ref{eq:delta_keff} and highlights the local behavior of the effective wave vector in distorted fields. Such features — including the so-called “extra recoil” effect~\cite{Bade2018}, where the momentum transfer exceeds the nominal value $h\nu/c$ — have remained inaccessible in atom interferometry experiments based on thermal clouds, due to spatial averaging over inhomogeneous fields.

To model our observations, we have developed two approaches. The first one is based on Monte-Carlo simulations:  we consider a set of classical trajectories representing the initial dispersion of position and velocity of the atoms. 

For each trajectory, we calculate the probability amplitude and the phase shift of the interferometer and then average over the cloud. To calculate the phase, we need to compute the phase of the lasers at the position where the photons are absorbed along the trajectory, as well as the associated recoil \cite{PSCT1994}. To do this, we numerically propagate a gaussian beam truncated by diaphragm to the position of the atoms using a Fourier transform. More details are given in the Methods section of \cite{morel_determination_2020}. The results of this simulation are represented by the dashed line in Fig.~\ref{Fig_donuts_1D}. This simulation shows very good agreement with the experimental data.

The second approach does not require the beam's analytical shape but only a measurement of the intensity profile as perceived by the atoms.

Indeed, in this situation, the contribution to the recoil correction due to the phase gradient is negligible and the main effect comes from the Laplacian of the intensity term (see \autoref{eq:delta_keff}) which can be evaluated from a map of the intensity. 

\begin{figure}
\includegraphics[width=1\columnwidth]{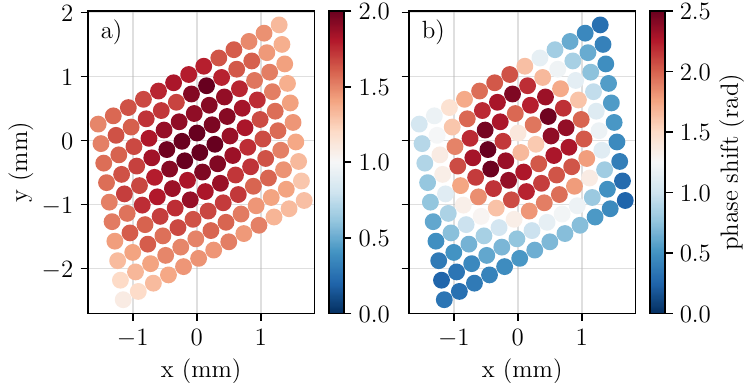}
\caption{
Reconstruction of beam intensity profile within the vacuum chamber using the light shift method with the BEC. Each data point corresponds to the phase shift value in radian for a position in the transverse plan, which is directly proportional to the intensity. (a) Spatial profile of the upward-propagating Bloch beam (b) spatial profile of the upward-propagating Bloch beam diffracted by an iris with an aperture of 4 mm diameter.
}
\label{fig:Fig_Light_shift}
\end{figure}

To map the local intensity profile in 2D we use a modified measurement sequence that consists of Ramsey interferometry sequences of two co-propagating Raman \(\pi/2\) pulses separated by 100 ms where an additional laser pulse - derived from the lower Bloch beam - is inserted in between. This pulse, whose power is stabilized by an active feedback system, induces a differential energy shift between the hyperfine states proportional to the laser intensity, allowing us to reconstruct the transverse intensity profile. Measurements are repeated while displacing the BEC across a fixed grid that covers a \(2\,\mathrm{mm} \times 2\,\mathrm{mm}\) area in the transverse plane. Fig.~\ref{fig:Fig_Light_shift} shows the intensity profiles extracted from these measurements of phase shifts induced by the upward propagating Bloch beam, both without and with iris diffraction. Due to the non-perpendicular orientation of the trapping beams, this grid is not orthogonal. In Fig.~\ref{fig:Fig_Light_shift}(b), the resulting profile exhibits a doughnut-shaped structure, in agreement with the propagation of a Gaussian beam truncated by a circular aperture.
Estimation of the recoil correction from the intensity is represented as red dots in Fig.~\ref{Fig_donuts_1D}. For each point, we recorded the intensity on a 3x3 matrix of adjacent points and calculated the Laplacian of the intensity to get the recoil correction. This theoretical model quantitatively reproduces the observed sign reversal, amplitude modulation, and spatial features in the measured \(\kappa_z\). This agreement supports the validity of the theoretical description and underscores the precision of our recoil mapping technique in resolving local optical distortions. The use of the Laplacian term is only an approximation and valid in the case of the small fluctuation or in this situation where we have mainly intensity fluctuations. Note that the doughnut intensity shape that we have here is similar to the intensity shape of a first order Laguerre-Gauss mode - however in that case the transverse phase gradient is not negligible \cite{barnett2013, afanasev2022} and in fact compensates the Laplacian term so that the longitudinal recoil does not exhibit extra recoil. It could be interesting to use our method to study effects related to wavefront topologies.

In this Letter, we present a robust method for directly measuring the spatial distribution of wave vectors and the intensity profile of laser beams, as experienced by atoms inside the vacuum chamber. Using atom interferometry techniques with a BEC, we locally measured both the photon recoil and the light shift induced by a probe laser. 
 
We have developed two theoretical approaches to estimate the local distribution of k-vectors, which show excellent agreement with the experimental data. This method provides a reliable tool for precisely characterizing systematic effects arising from imperfect wavefront profiles in atomic interferometers — effects that currently limit the ultimate sensitivity and accuracy of these devices. 
Our measurements also reveal a direct observation of an \text{extra recoil} exceeding the nominal value $h\nu/c$, which we deliberately enhanced by shaping the laser intensity profile.

At present, the statistical uncertainty of our measurements is mainly limited by the time required to scan the full beam profile. A promising solution to this limitation is to use 2D matter-wave arrays \cite{ArraysBEC2024} combined with advanced imaging techniques. A 2D matter-wave array would enable us to probe a larger portion of the beam's cross-section and to perform measurements at several transverse positions simultaneously, thereby significantly reducing the impact of temporal intensity fluctuations and long-term experimental drift. A recent work demonstrates in situ wavefront curvature characterization in a Mach-Zehnder ($\pi/2-\pi-\pi/2$) atom interferometer using an optical molasses \cite{junca2025}. It would be interesting to compare this approach with the one presented in this paper, in particular for evaluating the systematic effect caused by wavefront distortions.

\vspace{5mm}

This work was supported by the Agence Nationale pour la Recherche, TONICS Project No. ANR-21-CE47-0017, the DIM-Quantip, PEPR Quantique Project QAFCA (ANR- 22-PETQ-0005) and doctoral program of QuantEdu-France (ANR-22-CMAS-0001) in the framework of France 2030.

\bibliography{references}

\end{document}